\documentclass[12pt,preprint]{aastex}

\received{October 18, 2004}
\accepted{}

\def\sun{$_{\odot}$}
\def\um {$\mu$m}
\def\as {{$^{\prime\prime}$}}
\def\am {{$^\prime$}}
\def\deg{{$^\circ$}}

\begin{document}

\title{Quiescent Dense Gas in Protostellar Clusters: \\ 
the Ophiuchus A Core}

\author{James Di Francesco\altaffilmark{1}}
\affil{National Research Council of Canada, Herzberg Institute of Astrophysics, \\
5071 West Saanich Road, Victoria, BC V9E 2E7, Canada}

\author{Philippe Andr\'e}
\affil{Service d'Astrophysique, CEA/DSM/DAPNIA, \\
C.E. Saclay, 91191, Gif-sur-Yvette Cedex, France}

\author{Philip C. Myers} 
\affil{Harvard-Smithsonian Center for Astrophysics \\
60 Garden St. MS-42, Cambridge, MA, 02138-1516, U.S.A.}

\altaffiltext{1}{formerly at Radio Astronomy Laboratory, University of 
California, Berkeley, 601 Campbell Hall, Berkeley, CA, 94720-3411, U.S.A.}

\begin{abstract}
We present combined BIMA interferometer and IRAM 30 m Telescope data of 
N$_{2}$H$^{+}$ 1--0 line emission across the nearby dense, star forming core 
Ophiuchus A (Oph A) at high linear resolution (e.g., $\sim$1000 AU).  Six 
maxima of integrated line intensity are detected which we designate Oph A-N1 
through N6.  The N4 and N5 maxima are coincident with the starless continuum 
objects SM1 and SM2 respectively but the other maxima are not coincident with 
previously-identified objects.  In contrast, relatively little N$_{2}$H$^{+}$ 
1--0 emission is coincident with the starless object SM2 and the Class 0 
protostar VLA 1623.  The FWHM of the N$_{2}$H$^{+}$ 1--0 line, $\Delta V$, 
varies by a factor of $\sim$5 across Oph A.  Values of $\Delta V$ $<$ 0.3 km 
s$^{-1}$ are found in 14 locations in Oph A, but only that associated with 
N6 is both well-defined spatially and larger than the beam size.  Centroid 
velocities of the line, $V_{LSR}$, vary relatively little, having an rms of 
only $\sim$0.17 km s$^{-1}$.  Small-scale $V_{LSR}$ gradients of $<$0.5 km 
s$^{-1}$ over $\sim$0.01 pc are found near SM1, SM1N, and SM2, but not N6.  
The low N$_{2}$H$^{+}$ abundances of SM2 or VLA 1623 relative to SM1, SM1N, 
or N6 may reflect relatively greater amounts of N$_{2}$ adsorption onto dust 
grains in their colder and probably denser interiors.  The low $\Delta V$ of 
N6, i.e., 0.193 km s$^{-1}$ FWHM, is only marginally larger than the FWHM 
expected from thermal motions alone, suggesting turbulent motions in the Oph 
A core have been reduced dramatically at this location.  The non-detection of 
N6 in previous thermal continuum maps suggests that interesting sites possibly
related to star formation may be overlooked in such data.

\end{abstract}


\clearpage
\section{Introduction}

The numbers of embedded clusters observed in giant molecular clouds and 
the inferred birthrates of their stellar populations suggest strongly that 
stars in the Galaxy form predominantly within clustered environments and 
not in isolation (Lada \& Lada 2003).  At present, however, no paradigm 
exists describing how stellar formation proceeds within clusters inside 
turbulent cloud cores (Meyer et al. 2000).  Such cores can contain starless 
objects, i.e., local maxima of arbitrarily high column density where no 
evidence for star formation has been detected, as well as extremely young 
protostellar objects, i.e., locations where evidence for star formation has 
been detected.  Observations of these objects in clusters should significantly 
improve our understanding of how they form.  The gas and dust comprising or 
surrounding these objects can reveal the motions and physical conditions 
associated with the earliest stages of stellar formation, providing the 
necessary constraints to new models.

The Ophiuchus star-forming molecular cloud is particularly well suited for 
an intensive study of the dense gas associated with star formation within 
clusters.  Only 125 pc distant (de Geus, de Zeeuw, \& Lub 1989, de Geus 1992), 
Ophiuchus contains the nearest embedded cluster to the Sun (see Allen et al. 
2002, and references therein), residing within high column densities of gas 
and dust (e.g., see Loren, Wootten, \& Wilking 1990; LWW90).  Therefore, 
appropriate observations of Ophiuchus can reveal starless objects  and 
extremely young protostellar objects within a clustered environment at the 
best possible sensitivities and the finest possible linear resolutions.

Over the last decade, observations of the cores associated with L1688 in 
Ophiuchus, i.e., Oph A to Oph F (see LWW90), have revealed numerous potential 
starless and extremely young protostellar objects.  In particular, the Oph A 
core contains a protostellar cluster including the prototypical ``Class 0" 
object, VLA 1623, in addition to starless objects to its north and east named 
SM1N, SM1, and SM2 (see Andr\'e, Ward-Thompson, \& Barsony 1993; AWB93).  
Wide-field 1.3 mm continuum maps by Motte, Andr\'e, \& Neri (1998; MAN98) 
revealed Oph A is the brightest of the Oph cores, and a multiresolution 
wavelet analysis of its emission suggested the presence of eight other 
starless objects in Oph A.  The objects A-MM6, A-MM7, and A-MM8, together 
with SM1N, SM1, and SM2, comprise an arc within the filamentary structure 
of Oph A centered on the nearby B star S1 (see Figure 2a of MAN98).  MAN98 
also found the starless objects in the Oph cores have a mass spectrum 
reminiscent of the stellar IMF, suggesting a link exists between the 
formation of these objects and actual stars in clusters.

In this paper, we examine emission from dense gas associated with the Oph 
A core, using combined single-dish and interferometer data to attain a 
linear resolution of $\sim$1000 AU.  We observed the 1--0 transition of the 
N$_{2}$H$^{+}$ molecular ion since it is a tracer of dense gas (e.g., its 
critical density for excitation is $n_{cr}$ = 2 $\times$ 10$^{5}$ cm$^{-3}$) 
and N$_{2}$H$^{+}$ may deplete significantly only at densities $\>>$10$^{5}$ 
cm$^{-3}$ (see Bergin \& Langer 1997, Bergin et al. 2002).  Only two of the 
five maxima of integrated line intensity found are coincident with peaks of 
millimeter continuum emission from dust.  In addition, several peaks of dust
emission are not coincident with maxima of integrated line intensity.  This
situation is different from that seen in isolated cores where coincidence 
between dust and N$_{2}$H$^{+}$ emission is common.  The difference is likely 
due to variations in N$_{2}$H$^{+}$ abundance within Oph A arising from 
differences in the relative density and dust temperature within each object.  
Focusing on the line emission alone, we find that its widths vary by a factor 
of $\sim$5 in Oph A, with several locations having FWHMs $<$ 0.3 km s$^{-1}$.  
The centroid velocities of the line emission vary little in Oph A, however,
having an rms of only $\sim$0.17 km s$^{-1}$.  Small-scale centroid velocity
gradients are found near the SM1, SM1N, and SM2 objects.  The quiescent and 
slow motions of objects in Oph A objects contrast sharply with the turblent 
and dynamic motions expected in some models of cluster formation.  One 
integrated line intensity maximum, Oph A-N6, has line widths as low as 
0.193 km s$^{-1}$ FWHM, or just larger than the line width expected from 
thermal motions alone, indicating localized reduction of turbulent motions 
within Oph A.

Details of our observations are described in \S 2.  We present the results 
in \S 3, including derivations of N$_{2}$H$^{+}$ column densities and H$_{2}$ 
column densities, and descriptions of the widths, centroids, and profiles of 
the N$_{2}$H$^{+}$ 1--0 line across Oph A.   A discussion of the implications
of our data is found in \S 4, and we summarize our conclusions in \S 5.

\section{Observations}

\subsection{IRAM 30 m Telescope Observations}

Observations of the Oph A core at the 93.17 GHz frequency of N$_2$H$^+$ 1--0,
consisting of 7 hyperfine components over 4.6 MHz (see Caselli, Myers, \& 
Thaddeus 1995; CMT95), were carried out with the IRAM 30 m telescope on Pico 
Veleta, Spain\footnote{The Institut de RadioAstronomie Millim\'etrique (IRAM) 
is an international institute for research in millimeter astronomy, and is 
supported by the CNRS (Centre National de la Recherche Scientifique, France), 
the MPG (Max Planck Gesellschaft, Germany), and the IGN (Instituto Geografico 
Nacional, Spain).}.  Data were obtained during three clear nights between 1998 
June 25 and 1998 June 28, in association with a comprehensive molecular line 
study of the cores and condensations of the Ophiuchus protostellar cluster 
(Belloche et al.  2001; Andr\'e et al. 2004, in preparation)  Data were taken 
in the ``on-the-fly'' mapping mode (Ungerechts et al.  2000), with an 
individual dump time of 2 seconds, a scanning speed of 1\as\ s$^{-1}$, and a 
spacing of 5\as\ between consecutive scans.  The reference position was taken 
at the location of the B3 star S1 (see Figure 1) which had previously been 
seen to be free of emission in any dense gas tracer such as N$_{2}$H$^{+}$ 
1--0.  The half-power beam width of the telescope was $\sim$26\as.  An SIS 
heterodyne receiver was used with an autocorrelator as backend to give a 
spectral resolution of 19.5 kHz over 17.5 MHz.  The corresponding spectral 
resolution was 0.06284 km s$^{-1}$.

The observations were done in single sideband mode, with a sideband rejection 
of 0.003 and an associated calibration uncertainty of $\sim$10\%.  The forward 
and beam efficiencies of the telescope used to convert antenna temperatures 
$T^*_\mathrm{A}$ into main beam temperatures $T_\mathrm{mb}$ were 92\% and 
73\%, respectively.  The system temperature ranged from $\sim$170 K to 
$\sim$200 K on the  $T^*_\mathrm{A}$ scale.  The telescope pointing was 
checked every $\sim$2 hours on the strong radio sources NRAO 530 or 1514-241 
and found to be accurate to better than $\sim$5\as\ (i.e, the maximum deviation 
between two consecutive pointings).  The telescope focus, optimized on the 
same sources every $\sim$3-4 hours, was very stable.  The data cube was 
produced and reduced with the CLASS software package (Buisson et al. 2002).

\subsection{BIMA Millimeter Array Observations}
 
The Oph A core was observed with the BIMA Millimeter Array located at Hat 
Creek, California, U.S.A.\footnote{The BIMA array is operated with support 
from the U.S. National Science Foundation under grants AST-9981308 to the
University of California, Berkeley, AST-9981363 to the University of Illinois, 
and AST-9981289 to the University of Maryland.}, over 2 tracks in its 
C-configuration (1999 May 04, 23) and 2 tracks in its B-configuration (2000 
March 14, 15).  A total of 3 pointings was observed along the bright continuum 
filament associated with the A core detected by MAN98.  The pointings were 
spaced by $\sim$1\farcm0, slightly less than required for Nyquist sampling, 
for a constant rms noise level along the filament.  Each track consisted of 
cycled visits to the Oph A pointings interspersed with observations of radio 
sources 1625-254 (in 1999) or 1733-130 (in 2000) approximately every 20 
minutes for phase calibration.  Flux calibration was obtained from 
observations of 3C 273 and Neptune at the beginning and the end of each track 
respectively.  

Data reduction was performed using standard tasks in the MIRIAD software 
package (see Sault, Teuben \& Wright 1995).  Quasar data were self-calibrated 
using fluxes bootstrapped from Neptune data, and the resulting amplitude and 
phase solutions were applied to the target pointings.  All 3 pointings were 
inverted simultaneously, and the resulting mosaics were deconvolved with a 
cleaning algorithm based on that presented by Steer, Dewdney, \& Ito (1984).  
The N$_{2}$H$^{+}$ 1--0 data were convolved with a Gaussian of 5\farcs0 
$\times$ 0\farcs5 FWHM (P.A.  = 90\deg) to improve the beam shape and increase 
brightness sensitivity.  The final synthesized beam size was 9\farcs9 $\times$ 
6\farcs1 FWHM (and P.A. = -13.18\deg), corresponding to a linear FWHM 
resolution of 0.0060 pc $\times$ 0.0037 pc at the 125 pc distance to Oph A.

\subsection{Combining the Datasets}

The BIMA and IRAM data of N$_{2}$H$^{+}$ 1--0 were combined to recover 
missing flux from extended structure resolved out in the interferometer map.  
The IRAM data were regridded in velocity from 0.06284 km s$^{-1}$ channels 
to 0.07854 km s$^{-1}$ channels to match the velocity resolution of the BIMA 
data.  The MIRIAD task {\it immerge} was used for the combination, where the 
sum of the Fourier transforms of the clean BIMA and IRAM images is itself 
transformed into the image plane as the combined image.  To ensure a common 
flux scale for the independent datasets prior to combination, visibilities 
of both were compared over 2--9 k$\lambda$, the maximum range of spatial 
frequencies in common, and a derived scaling factor of 1.11 was applied to 
the IRAM data.  Comparing the total flux of the combined dataset convolved 
to a $\sim$27\as\ FWHM beam with that of the IRAM data, we estimate this 
method recovers 85\% of the flux in the latter map over the region of interest.

\section{Results}

Figure 1a shows the 1.3 mm continuum emission from Oph A detected by MAN98 
using the IRAM 30 m Telescope at 13\as\ FWHM resolution.  Figure 1b shows 
the integrated intensity of N$_{2}$H$^{+}$ 1--0 line emission detected by 
us over the same region, using combined data from the IRAM 30 m Telescope 
and the BIMA millimeter interferometer at $\sim$10\as\ $\times$ $\sim$6\as\ 
FHWM resolution.  The line emission from Oph A shows a filamentary structure 
similar to that shown by the continuum emission, although the width of the
the integrated line emission structure is narrower than that of the continuum 
emission.  Figure 1a shows positions of several objects in Oph A identified 
by MAN98 from their multiresolution wavelet analysis.  Figure 1b shows these 
positions again as well as the positions of the various maxima of integrated 
intensity of N$_{2}$H$^{+}$ 1--0, which we have named here Oph A-N1 through 
N6 and whose positions are listed in Table 1.  Some, but not complete, 
coincidence is seen between continuum objects and the maxima of integrated 
line intensity in Oph A.  For example, SM1 and SM1N are coincident (within 
a beam FWHM) of N4 and N5 respectively (and possibly A-MM5 with N2 also).  
Some Oph A continuum objects, however, have little-to-no detected 
N$_{2}$H$^{+}$ 1--0 emission associated with them.  For example, relatively 
little line emission is found at the positions of VLA 1623, SM2, A-MM6, A-MM7, 
and A-MM8.  Furthermore, some maxima of the N$_{2}$H$^{+}$ 1--0 integrated 
intensity in Oph A, although coincident with extended continuum emission, 
are not associated with continuum objects.  For example, N1, N3, and N6 are 
not associated with objects specifically identified by MAN98.  The lack 
of coincidence between peaks of dust emission and maxima of N$_{2}$H$^{+}$ 
emission in Oph A is very different from the high level of coincidence 
between the two tracers seen in numerous isolated cores, albeit at lower 
resolution (e.g., Tafalla et al.  2002).

All 7 hyperfine components of the N$_{2}$H$^{+}$ 1--0 line from the combined 
data were fit simultaneously using the HFS (HyperFine Structure) fitting 
routine of the CLASS data analysis package (see Buisson et al.)  For each 
0\farcs75 $\times$ 0\farcs75 pixel of the data cube, these fits provided 
estimates of: 1) $\Delta V$, the line FWHM, 2) $V_{LSR}$, the centroid 
velocity, 3) $\tau_{tot}$, the sum of peak optical depths for all 7 hyperfine
components, and 4) ($J_{ex}$--$J_{bg}$)$\tau_{tot}$, the difference between 
the (Rayleigh-Jeans equivalent) excitation and background temperatures times
$\tau_{tot}$.  (In the latter, $J_{ex}$ = ($h\nu/k$)/(exp($h\nu/kT_{ex}$)-1), 
$J_{bg}$ = ($h\nu/k$)/(exp($h\nu/kT_{bg}$)-1), and $T_{ex}$ and $T_{bg}$ 
are the actual excitation and background temperatures respectively.)  The 
brightness temperature of the 1--0 line, $T_{B}$, can be estimated from 

$$ T_{B} = (J_{ex}-J_{bg})(1-{\rm exp}(-\tau_{tot})). \eqno{(1)} $$

\noindent
Table 2 lists the mean, rms, minimum and maximum values of $\Delta V$, 
$V_{LSR}$, $T_{B}$ and $\tau_{tot}$ from the combined data after averaging 
them into 10\farcs5 $\times$ 6\farcs75 pixels, i.e., pixels with areas 
slightly larger than those of the synthesized beam.  Most notably, $\Delta 
V$ in Oph A varies widely but $V_{LSR}$ does not.  For example, $\Delta V$ 
varies by a factor of $\sim$5 in Oph A but the rms of $V_{LSR}$ is only 0.17 
km s$^{-1}$. 

In the following, we describe the N$_{2}$H$^{+}$ data in more detail, using 
the results obtained by HFS fitting to examine the N$_{2}$H$^{+}$ column 
densities and abundances as well as the widths and centroid velocities of 
the N$_{2}$H$^{+}$ 1--0 line.

\subsection{Column Densities and Abundances}

Column densities of H$_{2}$, i.e., $N$(H$_{2}$), can be determined in Oph
A with the 1.3 mm continuum data of MAN98 (e.g., see Figure 1a) from
$$ \it{N}\rm{(H_{2})} = \it{S}_{\nu}/(\rm{\Omega_{m}} \mu {\it m_{H}} \kappa_{\nu}\rm{B}_{\nu}({\it T_{dust}})), \eqno{(2)} $$

\noindent
where $S_{\nu}$ is the 1.3 mm flux density, $\Omega_{m}$ is the main-beam 
solid angle, $\mu$ is the mean molecular weight, $m_{H}$ is the mass of 
atomic hydrogen, $\kappa_{\nu}$ is the dust opacity per unit mass at 1.3 mm 
(assumed to be 0.005 cm$^{2}$ g$^{-1}$ for starless objects by MAN98; 
the local gas-to-dust ratio is assumed to be 100), and B$_{\nu}$($T_{dust}$) 
is the Planck function at T = $T_{dust}$.

Figure 2a shows $N$(H$_{2}$) in Oph A obtained by assuming $T_{dust}$ = 20 
K at all locations (following MAN98).  To facilitate comparision with the 
N$_{2}$H$^{+}$ data, the area shown is restricted to that where the peak
brightness temperature of the brightest hyperfine component, $T_{B}^{max}$ 
$\geq$ 3 $\sigma$\/ in the N$_{2}$H$^{+}$ 1--0 line data.  Furthermore, the 
MAN98 continuum data were binned into pixels 10\farcs5 $\times$ 6\farcs75 in 
size using standard image interpolation methods in MIRIAD to obtain pixels 
with areas approximately that of the beam of the N$_{2}$H$^{+}$ observations.  
(Note that each pixel in Figure 2a is still not spatially independent of 
its neighbors.)  Figure 2a shows the $N$(H$_{2}$) distribution across Oph 
A is similar to that of the 1.3 mm continuum emission shown in Figure 1a, 
due of course to the linear proportionality between the two quantities in 
Equation 2 if $T_{dust}$ is constant.  Table 3 lists the mean, rms, minimum, 
and maximum of $N$(H$_{2}$) shown in Figure 2a. 
 
Column densities of N$_{2}$H$^{+}$, i.e., $N$(N$_{2}$H$^{+}$), can be 
determined along the Oph A filament from the line characteristics obtained 
from HFS fitting (see \S 3).  Assuming LTE, i.e., all level populations are 
described by a single excitation temperature, $T_{ex}$, $N$(N$_{2}$H$^{+}$) 
can be found following Womack, Ziurys, \& Wyckoff (1992) from

$$ {\it N}({\rm N_{2}H^{+}}) \approx {1\over{1.06(J+1)}}{3h\over{8\pi^{3}\mu_{e}^{2}}}{k\over{hB}}\Biggl({T_{ex}+{hB\over{3k}}\Biggr)}{e^{E_{J}/kT_{ex}}\over{(1-e^{-h\nu/kT_{ex}})}}{\Delta V}{\tau_{tot}}, \eqno{(3)} $$

\noindent
where $\Delta V$ and $\tau_{tot}$ are found from HFS fits to data averaged to 
10\farcs5 $\times$ 6\farcs75 pixels.  In addition, $J$ is the lower rotational 
level number (i.e., 0), $\mu_{e}$ is the dipole moment of the N$_{2}$H$^{+}$ 
molecular ion (i.e., 3.40 Debye; Green, Montgomery \& Thaddeus 1974), $B$ is 
the rotational constant of N$_{2}$H$^{+}$ (i.e., 46.586702 GHz; see CMT95) and 
$E_{J}$ is the energy above ground of the lower rotational level (i.e., 0).
%
%

Figure 2b shows $N$(N$_{2}$H$^{+}$) in Oph A obtained using the results 
of our HFS fits and Equation 3.\footnote{Using Equation 3, $\sim$80\% of 
beam-sized pixels had derived values of $T_{ex}$ between 6 K and 24 K with a 
peak at $\sim$15 K, but the remaining $\sim$20\% had values extending up to 
260 K.  These latter values were found typically if $\tau_{tot}$ $<$ 0.2;
high values of $T_{ex}$ are obtained in such cases because only the product 
of $J_{ex}$ and $\tau_{tot}$ is constrained.  In these instances, the derived 
$T_{ex}$ values were deemed spurious, and $N$(N$_{2}$H$^{+}$) values were 
calculated using Equation 3 after substituting $T_{B}$ for the factor 
($T_{ex}$ + $hB/3k$)$\tau_{tot}$ and assuming in the exponential factors 
$T_{ex}$ = 14.9 K, the mean $T_{ex}$ of the other $\sim$80\% of pixels.}  
Figure 2b shows the $N$(N$_{2}$H$^{+}$) distribution across Oph A is not 
similar to the $N$(H$_{2}$) distribution, but rather is similar to that of 
the N$_{2}$H$^{+}$ integrated intensities shown in Figure 1b.  Table 3 lists 
the mean, rms, minimum, and maximum of $N$(N$_{2}$H$^{+}$) shown in Figure 
2b.  We note that assuming LTE in Oph A can result in overestimates of 
$N$(N$_{2}$H$^{+}$) from N$_{2}$H$^{+}$ 1--0 emission toward locations with 
densities $<$ 10$^{6}$ cm$^{-3}$ (e.g., by factors of $\sim$2 where densities
are $\sim$10$^{4}$ cm$^{-3}$; E. Bergin, private communication.)  As discussed 
in \S 4.1, however, mean densities are $>$10$^{6}$ cm$^{-3}$ towards positions 
of bright N$_{2}$H$^{+}$ 1--0 emission in Oph A.

Figure 2c shows the N$_{2}$H$^{+}$ fractional abundance across Oph A, 
determined using values from Figures 3a and 3b (i.e., $X$(N$_{2}$H$^{+}$) = 
$N$(N$_{2}$H$^{+}$)/$N$(H$_{2}$)).  The mean $X$(N$_{2}$H$^{+}$) derived, 1.9
$\times$ 10$^{-10}$, is similar to values derived for other dark clouds, e.g., 
$X$(N$_{2}$H$^{+}$) $\approx$ 10$^{-10}$ has been found in various cores by 
Caselli et al. (2002a), Tafalla et al. (2002), Benson, Caselli, \& Myers 
(1998), Womack et al. (1992), and Ohishi, Irvine, \& Kaifu (1992).  Within 
Oph A itself, Figure 2c shows that $X$(N$_{2}$H$^{+}$) varies by a factor of 
$\sim$40, with lower values at the edges of the filament and near the 
continuum objects like SM2, and higher values at positions at the extreme 
northern and southern ends of the filament.  Table 3 lists the mean, rms, 
minimum, and maximum of $X$(N$_{2}$H$^{+}$) shown in Figure 2c.  

Assuming LTE, our accuracy in determining $X$(N$_{2}$H$^{+}$) is limited 
by our knowledge of $T_{dust}$ within Oph A\footnote{In addition, dust mass 
opacities, $\kappa_{\nu}$, may vary in Oph A by factors of $\sim$2 or more 
due to grain evolution (see MAN98 and references therein).  In this paper, 
we assume constant $\kappa_{\nu}$ since we are considering similar objects 
where differences in dust opacities are likely minimal.}.  Indeed, isothermal 
dust is arguably not expected in Oph A given the number of nearby heating 
sources (i.e., embedded cluster members) and the wide range of column densities 
(i.e., extinctions) present.  For the continuum objects, dust temperatures 
determined independently can be used to determine more refined, specific 
values of $X$(N$_{2}$H$^{+}$).  Table 4 lists values of $T_{dust}$ for 
continuum objects in Oph A by AWB93, determined by modeling their 
submillimeter spectral energy distributions.  (AWB93 found dust emissivities 
of $\kappa_{\nu}$ $\propto$ $\nu^{1.5}$, i.e., $\kappa_{1.3 mm}$ $\approx$ 
0.01 cm$^{2}$ g$^{-1}$, best fit their data.)  AWB93 speculated these 
differences may be due to radiative heating by nearby sources to the 
northeast of SM1, e.g., the nearby B3 star S1.  Table 4 also lists values 
of $X$(N$_{2}$H$^{+}$), $N$(N$_{2}$H$^{+}$), and $N$(H$_{2}$) for VLA 1623, 
SM1N, SM1, and SM2 obtained by assuming the $T_{dust}$ values of AWB93 in 
Equations 2 and 3 and averaging the resulting values over 3-4 beam-sized 
pixels coincident with each object.  (For VLA 1623, upper limits for 
$N$(N$_{2}$H$^{+}$) and $X$(N$_{2}$H$^{+}$) are given in Table 4 since 
it is located where $T_{B}^{max}$ $<$ 3 $\sigma$.)  The difference in 
$X$(N$_{2}$H$^{+}$) from one source to the next is smaller when constant 
$T_{dust}$ is used.  N$_{2}$H$^{+}$ abundances are similar for SM1N and 
SM1 while those of SM2 and VLA 1623 are lower by factors $>$4.  (Oph A-N6, 
also in Table 4, will be discussed in \S 3.2.)

For locations in Oph A beyond the continuum objects, specific values of 
$T_{dust}$ have not been defined, e.g., AWB93 did not derive $T_{dust}$ 
values there.  Maps of submillimeter continuum spectral indicies by AWB93,
however, indicate that a $T_{dust}$ gradient exists across Oph A, with 
lower $T_{dust}$ (i.e., $\leq$ 15 K) to the south or southwest (near VLA 
1623 and SM2) and higher $T_{dust}$ (i.e., $\geq$ 25 K) to the north or
northeast (near SM1 and SM1N).  Unfortunately, the AWB93 submillimeter
(e.g., 350 \um) continuum maps do not cover well the extreme northern 
and southern ends of the filament where we find the highest values of 
$X$(N$_{2}$H$^{+}$), i.e., assuming $T_{dust}$ = 20 K.  Without specific 
estimates for $T_{dust}$, it is difficult to refine our estimates of 
$X$(N$_{2}$H$^{+}$) at the ends of the filament.  We note, however, that 
$X$(N$_{2}$H$^{+}$) at these locations would be quite similar to those of 
SM1 and SM1N if $T_{dust}$ $\approx$ 10--12 K, quite significantly reducing 
the abundance contrast shown by Figure 2c.  At the very least, our 
observations suggest differences in $X$(N$_{2}$H$^{+}$) for the continuum 
objects in Oph A by factors $\geq$4.  Put simply, variations of both 
$X$(N$_{2}$H$^{+}$) and $T_{dust}$ appear necessary to reconcile the 1.3 
mm continuum and N$_{2}$H$^{+}$ line data of Oph A.  

\subsection{Line Widths}

Figure 3a shows a map of $\Delta V$ obtained from the HFS fits of the 
combined N$_{2}$H$^{+}$ data of Oph A.  The data shown are only those 
where $T_{B}^{max}$ $\geq$ 3 $\sigma$.  The large variation of $\Delta 
V$ in Oph A by a factor of $\sim$5 differs from the relatively little
variation of line widths (i.e., ``coherence") found over similar $\sim$0.1 
pc length scales in isolated, quiescent cores by Goodman et al. (1998).  
Table 4 lists the $\Delta V$ of SM1N, SM1, and SM2, obtained by averaging 
over 4 beam-sized pixels surrounding each object.  These continuum objects 
are coincident with locations in Oph A with relatively broad N$_{2}$H$^{+}$ 
lines, i.e., $\Delta V \approx$ 0.5--0.6 km s$^{-1}$.  We note also that 
the largest line widths are found adjacent to (but not coincident with)
these continuum objects.

Figure 3a shows fourteen locations in Oph A with $\Delta V$ $\leq$ 0.30 km 
s$^{-1}$, which we arbitrarily define as ``very narrow."  Table 5 lists these
positions, named here as locations ``a" through "n", and gives their mean line
width within the 0.3 km s$^{-1}$ contour.  No very narrow line width location 
is coincident with an identified continuum object, although location ``c", 
near A-MM5, may be an exception.  Eleven very narrow line width locations are 
at the edges of the detected filament, where line peaks are relatively weak 
and near our detection limit (i.e., $T_{B}^{max} \approx 3$ $\sigma$) and 
where HFS fits are relatively poor (e.g., typical $\Delta V$ uncertainties 
are $\sim$0.02-0.05 km s$^{-1}$.)  Conversely, only the ``b", ``f, and ``l" 
locations are well-defined local minima of $\Delta V$; each is surrounded by 
larger $\Delta V$ and has line peaks well above our detection limit (e.g., 
typical $\Delta V$ uncertainties are $\sim$0.005-0.01 km s$^{-1}$).  The 
extents of most very narrow line locations are therefore not well defined 
spatially since narrow line emission below our detection limit may or may 
not reside adjacent to each.  Finally, most narrow line locations have 
detected extents smaller than the beam.  Only the ``c", ``d", ``l", and ``m" 
locations have detected extents larger than the beam.

The very narrow line location ``l" is especially notable in Oph A given 
its resolved extent, the relative brightness and the narrowness of its 
N$_{2}$H$^{+}$ 1--0 line emission, and its association with a maximum of 
integrated intensity (i.e., Oph A-N6).  The ``b" and ``f" locations are 
like ``l" in that they are also associated with bright emission (i.e., 
Oph A-N1), but unlike ``l" they are each spatially unresolved and their 
lines are slightly broader than those of ``l".  The ``c", ``d", and ``m" 
locations are like ``l" in that they are each spatially resolved and their 
lines are similarly as narrow than those of ``l" (or less), but unlike ``l" 
they are not spatially well-defined and have weaker emission (i.e., no 
associations with integrated intensity maxima).  These limitations make the 
``b", ``c", ``d", ``f", and ``m" locations more difficult to characterize 
than ``l", so we concentrate on ``l" in the following discussion.  (The 
conclusions we reach on ``l" may also apply to these other very narrow 
line positions.)  Since ``l" includes Oph A-N6 we henceforth refer to this 
``object" or location within Oph A simply as Oph A-N6 or N6 for short.  
Table 4 lists values of $X$(N$_{2}$H$^{+}$), etc., for N6 found by averaging 
values over 5 beam-sized pixels coincident with positions where $\Delta V$ 
$\leq$ 0.30 km s$^{-1}$, and assuming $T_{dust}$ = 15 K at that location 
(see \S 3.1).  Although the position of minimum line width in N6 is 
$\sim$19\as\ from its position of maximum integrated intensity, line 
emission is quite narrow throughout N6; when averaged over 5 adjacent 
beam-sized pixels, the $\Delta V$ of N6 is 0.25 km s$^{-1}$ (see Table 4).  

The minimum line width in Oph A-N6 is remarkably small, only 0.193 $\pm$ 
0.008 km s$^{-1}$.  Figure 4 shows the spectrum of N$_{2}$H$^{+}$ 1--0 at 
the location of minimum line width in Oph A-N6, along with its HFS fit.  
All 7 hyperfine components of N$_{2}$H$^{+}$ 1--0 are unblended and clearly 
discernable at this position.  Given the finite 0.07854 km s$^{-1}$ velocity 
resolution, the actual line widths are slightly less than the values shown 
in Figure 3a, depending on the line width.  For example, the minimum $\Delta 
V$ in Oph A-N6 reduces to only 0.176 km s$^{-1}$ when the velocity resolution
is subtracted in quadrature.  Such narrow line widths are actually comparable 
to the thermal FWHMs expected for N$_{2}$H$^{+}$ given expected gas 
temperatures, $T_{gas}$, within Oph A.  For example, Wootten et al. (2004, 
in preparation) using VLA NH$_{3}$ data have determined gas temperatures for 
the Oph A continuum objects that are similar to the $T$ = 20 K assumed by 
MAN98, i.e., 16 K for SM1, 21 K for SM1N, and 18 K for SM2.  Other estimates 
of $T_{gas}$ in Oph A can be made using the values of $T_{ex}$ found from 
HFS fits of the N$_{2}$H$^{+}$ 1--0 line at positions where $\tau_{tot}$ $>>$ 
1, assuming the transition is thermalized, i.e., the local density is well 
above its 2 $\times$ 10$^{5}$ cm$^{-3}$ critical density.  (Below we show 
that this is reasonable within Oph A-N6, given its size and mass.)  For the 
minimum $\Delta V$ spectrum shown in Figure 4, values of $T_{ex}$ = 18.1 K 
$\pm$ 3.4 K and $\tau_{tot}$ = 12.0 $\pm$ 2.2 were found, suggesting $T_{gas}$ 
$\approx$ 18 K.  Within Oph A-N6 where $\tau_{tot}$ $\geq$ 11.57 (i.e., where 
the optical depth of the brightest hyperfine component, 123-012, is at least 
3), the mean $T_{ex}$ is 17.4 K.  If $T_{gas}$ = 18.1 K at the position of 
the spectrum shown in Figure 4, the expected thermal FHWM is 0.169 km s$^{-1}$. 
The FWHM of the nonthermal component to $\Delta V$ at this position, $\Delta 
V_{NT}$, obtained by subtracting in quadrature the thermal FWHM, $\Delta 
V_{T}$, from the resolution-corrected FWHM, is only 0.0500 km s$^{-1}$.  
Although the spectrum shown in Figure 4 is the narrowest in Oph A-N6, the 
gas velocities there are all remarkably dominated by thermal motions, with 
relatively little nonthermal contributions.

The occurrence of near-thermal line widths in Oph A is significant because 
the Ophiuchus star-forming clouds are turbulent on larger spatial scales and
at lower densities (see Loren 1989a).  Near-thermal line widths have been 
observed previously towards isolated cores; e.g., see Jijina, Myers, \& Adams 
(1999) for a detailed list based on NH$_{3}$ observations and Benson et al. 
(1998) and Caselli et al. (2002a) for N$_{2}$H$^{+}$ observations.  Oph A-N6 
is similar to the quiescent regions in Serpens NW identified using 
N$_{2}$H$^{+}$ 1--0 by Williams \& Myers (2000).  For example, the average 
$\Delta V_{NT}$ for Oph A-N6, assuming $T_{gas}$ = 18 K, is 0.17 km s$^{-1}$, 
similar to objects at the lower end of the $\Delta V_{NT}$ range reported by 
Williams \& Myers.  There are, however, some important differences between 
Oph A-N6 and the quiescent Serpens objects.  For example, the minimum $\Delta 
V_{NT}$ of Oph A-N6 is smaller than those reported for the Serpens objects by 
a factor of $\sim$3.  Furthermore, Oph A-N6 contains an obvious maximum of 
N$_{2}$H$^{+}$ 1--0 integrated intensity, whereas the Serpens objects are 
not.  Other instances of N$_{2}$H$^{+}$ 1--0 as narrow as 0.2 km s$^{-1}$ 
were detected within NGC 1333 IRAS 4 by Di Francesco et al. 2001 (see also 
Walsh et al. 2004, in preparation, for other locations in NGC 1333.)  In 
contrast to Oph A-N6, $\Delta V_{NT}$ = 0.46-0.57 km s$^{-1}$ for SM1N, SM1, 
and SM2, given their respective values of $\Delta V$ listed in Table 4 and 
the $T_{gas}$ values found by Wootten et al.  

Within Ophiuchus, extremely narrow lines such as those described here have
been uncommonly observed, possibly due to the relatively large beam sizes of
earlier studies.  For example, Loren (1989b) found $\Delta V$ = 1.7 km s$^{-1}$ 
for $^{13}$CO 1--0 in the general Oph A region (R22) when observed with a 
2.4\am\ beam (although the line was considered saturated.)  In addition, 
Tachihara, Mizuno, \& Fukui (2000) found $\Delta V$ = 1.5 km s$^{-1}$ for the 
C$^{18}$O 1--0 core in the neighbourhood of Oph A ($\rho$ Oph 2).  Furthermore, 
LWW90 found $\Delta V$ = 0.8 km s$^{-1}$ and 0.7 km s$^{-1}$ for DCO$^{+}$ 
3--2 at the ``core center" of Oph A (defined then as $\sim$10\as\ west of SM1N) 
when observed with a 1\farcm2 FWHM beam and a 0\farcm5 FWHM beam respectively.  
Belloche et al. (2001), however, detected relatively narrow N$_{2}$H$^{+}$ 
1--0 lines at various locations within Oph, e.g., $\Delta V$ = 0.26 km s$^{-1}$ 
toward Oph E-MM2, in IRAM 30 m Telescope data alone (obtained with a 0\farcm45 
FWHM beam.)  Our IRAM 30 m Telescope data is a subset of the Belloche et al. 
data and across Oph A-N6 they show no evidence of the narrower lines seen 
in the combined data, i.e, $\Delta V$ = 0.35 km s$^{-1}$.  Our high angular 
resolution data have revealed locations of extremely narrow line emission 
in Oph A.

\subsection{Line Centroids and Profiles}

Figure 3b shows a map of $V_{LSR}$, the central velocity of the N$_{2}$H$^{+}$ 
1--0 line, in Oph A obtained from HFS fits to the combined data.  Immediately 
striking in this Figure is the narrow range of $V_{LSR}$ values within Oph A; 
the rms of $V_{LSR}$ values is only 0.17 km s$^{-1}$ (see Table 1).  Figure 
3b reveals no large-scale gradient of $V_{LSR}$ along Oph A but small-scale 
gradients are possibly found near the continuum objects SM1N, SM1, and SM2, 
roughly aligned east-west or west-east.  Other possible gradients include one 
north of VLA 1623 aligned north-south, and one northwest of SM1N (and Oph A-N1)
aligned northwest-southeast.  In contrast, no $V_{LSR}$ gradient is found 
toward Oph A-N6.  (As noted earlier, weak emission in its periphery are at 
slightly lower $V_{LSR}$.) 

Figure 5 shows a spectral grid of N$_{2}$H$^{+}$ 101--012, the ``isolated 
component" of N$_{2}$H$^{+}$ across Oph A, where each panel shows the line 
averaged over a (beam-sized) 10\farcs5 $\times$ 6\farcs75 area.  The variation 
of $\Delta V$ and lack of variation of $V_{LSR}$ across Oph A are easily 
apparent in this Figure, but the line shapes are more evident here than 
possible from maps of HFS fit parameters.  Figure 5 reveals that the 
N$_{2}$H$^{+}$ line profiles in Oph A are typically symmetric in most 
locations, with little evidence for self-reversal or asymmetry.  This 
appearance is consistent with the relatively low optical depths of the 
lines here as obtained from HFS fits (see Figure 2b).  Only at a position 
just west of SM1 is a single self-reversal found, and perhaps not 
coincidentally this location is also that of maximum $N$(N$_{2}$H$^{+}$) 
in Oph A as derived from the HFS fits.  The line shapes reveal little 
evidence for the asymmetrically-blue line profiles that suggest infall 
motions, such as those seen in N$_{2}$H$^{+}$ 1--0 towards L1544 by 
Williams et al. (1998), likely also due to the relatively low optical 
depths of the line in Oph A.  Only one possible instance of blue asymmetry 
is found in Oph A, immediately southwest of SM1.  This spectrum, however,
is adjacent to an instance of red asymmetry, and both asymmetries may be 
consistent with having been made from noise across the line.  The line 
shapes also show no evidence for broad line wings expected from high 
velocity outflows. 

Surprisingly, the line velocities of N$_{2}$H$^{+}$ 1--0 appear correlated
with its line width, but only for the brightest lines in Oph A, i.e. towards
the highest N$_{2}$H$^{+}$ column densities.  Figure 6a shows that higher 
velocity (redder) lines are systematically wider in lines where $T_{B}^{max}$ 
$\geq$ 10 $\sigma$.  A linear fit to the data shown in Figure 6a reveals a 
fairly high correlation parameter $r$ of 0.82 (see Press et al. 1992).  
Figure 6b shows, however, that a similar correlation between $V_{LSR}$ 
and $\Delta V$ is less likely for lines with $T_{B}^{max}$ $<$ 10 $\sigma$;
the $r$ obtained from a linear fit to this subsample is only 0.31.  Figure 
7 shows evidence that the line widths are increasing because only their red 
sides are moving to higher velocities and their blue sides are not shifting 
in velocity.  Three pairs of 2 spectra of the N$_{2}$H$^{+}$ 101-012 line 
are shown, taken from different positions in Oph A, where each pair consists 
of an average over 2-4 neighboring pixels with bright narrow emission in 
Oph A and an average over 4-6 surrounding pixels with fainter, broader 
emission.  This Figure suggests that the trend shown in Figure 6a is not 
simply a matter of the widths of bright lines increasing as the entire line 
shifts to higher velocities.  A similar trend is suggested by N$_{2}$H$^{+}$ 
1--0 line data of the quiescent region Q8 of Serpens (Williams \& Myers) 
and NGC 1333 IRAS 4B in Perseus (Di Francesco et al.)

\section{Discussion}

Our combined single-dish and interferometer data have revealed interesting
variations of N$_{2}$H$^{+}$ abundance and the width and centroid velocity
of the N$_{2}$H$^{+}$ 1--0 line within the Oph A filament.  Together, these 
variations provide a qualitative picture of the protostellar evolution of 
gas and dust in a relatively dense star-forming core, assuming that the
differences are due solely to evolution.  In this picture, the ``object" Oph 
A-N6 represents a relatively early, quiescent stage marked by high density, 
low $T_{dust}$, high $X$(N$_{2}$H$^{+}$), low $\Delta V$, and little local 
variation of $V_{LSR}$.  SM2, however, represents a more-evolved protostellar 
stage with higher density, low $T_{dust}$, low $X$(N$_{2}$H$^{+}$), medium 
$\Delta V$, and some local variation of $V_{LSR}$.  SM1N and SM1 may be at a 
similar protostellar stage as SM2 given their similar $\Delta V$ and variation 
of $V_{LSR}$, but their higher $X$(N$_{2}$H$^{+}$) may reflect their higher
$T_{dust}$.  Finally, VLA 1623, which has just formed a protostar (AWB93), 
is representative of the next stage of evolution, namely the beginning of 
the protostellar phase.  It has high density, low $T_{dust}$, and very low 
$X$(N$_{2}$H$^{+}$).  In the following, we describe in more detail how the 
differences in $X$(N$_{2}$H$^{+}$), $\Delta V$, and $V_{LSR}$ between the 
Oph A objects result in this qualitative picture.

\subsection{N$_{2}$H$^{+}$ Abundances}

Variations of N$_{2}$H$^{+}$ abundance in Oph A point to subtle differences 
in the evolutionary stages of its starless objects due to the expected
behavior of $X$(N$_{2}$H$^{+}$) within cold, dense environments.  Although 
$X$(N$_{2}$H$^{+}$) can be reduced locally as a result of outflows, e.g., when 
H$_{2}$O and CO are returned to the gas phase by shocks and quickly deplete 
N$_{2}$H$^{+}$ via ion-molecule reactions (see Bergin, Neufeld, \& Melnick 
1998; van Dishoeck \& Hogerheijde 1999), these objects do not contain or are 
not impacted by any outflowing gas (Kamazaki et al. 2001, 2003).  Lee et al. 
(2003) have suggested molecular abundance differences between starless 
objects can be related also to the relative ratio of their dynamical and 
chemical time scales (i.e., their respective environments), but the Oph A 
objects should have this factor in common since they are situated within the 
same core\footnote{Again, variations in grain size distribution may be relevant 
to abundance differences but we have assumed constant $\kappa_{\nu}$ throughout 
Oph A.}.  Gas-grain chemistry at high densities and low temperatures is a very
plausible cause of these variations.  

Examples of the effect of gas-grain chemistry on $X$(N$_{2}$H$^{+}$) in 
cold, dense environments were recently provided by Aikawa, Ohashi, \& Herbst 
(2003; AOH03). AOH03 found from isothermal (i.e., $T_{K}$ = 10 K) models of 
Larson-Penston collapsing starless objects that central $X$(N$_{2}$H$^{+}$) 
increases when central densities increase up to $\sim$3 $\times$ 10$^{6}$ 
cm$^{-3}$ (e.g., that expected for L1544), as the principal destroyers of 
N$_{2}$H$^{+}$ like CO become strongly depleted onto dust grains (see also 
Bergin \& Langer, Shematovich et al.\/ 2003).  At densities $\approx$ 3 
$\times$ 10$^{7}$ cm$^{-3}$, however, AOH03 found that $X$(N$_{2}$H$^{+}$) 
{\it decreases} by a factor of $\sim$2 from its maximum within a $\sim$1000 
AU radius as N$_{2}$, from which N$_{2}$H$^{+}$ forms, is adsorbed onto dust 
grains.  Although not explored explicitly by AOH03, further N$_{2}$H$^{+}$ 
depletion may be expected at still higher central densities.  Observational 
evidence for central N$_{2}$H$^{+}$ depletions have been reported recently 
in isolated starless objects with central densities of $\sim$10$^{5-6}$ 
cm$^{-3}$, e.g., B68 (Bergin et al.), L1544 (Caselli et al. 2003), and
IRAM 04191+1522 (Belloche \& Andre 2004, in preparation).

The models of AOH03 imply that depletion of N$_{2}$H$^{+}$ can be hard to 
observe directly with large beams.  For example, no central minimum in 
$N$(N$_{2}$H$^{+}$) can be seen when their model core is ``observed" with 
a beam of $\sim$4000 AU equivalent FWHM (see their Figure 9).  Given their 
small extent, such N$_{2}$H$^{+}$ depletion zones would be easier to detect 
directly with smaller beams.  For example, our combined N$_{2}$H$^{+}$ data 
have a geometrical mean linear resolution of 970 AU at the 125 pc distance 
of Oph A, or about half the diameter of the central N$_{2}$H$^{+}$ depletion 
zone in the AOH03 models.  The variations of $X$(N$_{2}$H$^{+}$) between the 
Oph A objects we observe could be then explained as the result of detecting 
differing amounts of depletion due to density differences between objects.  
MAN98 determined mean densities for SM1N, SM1, and SM2 from their 1.3 mm 
continuum observations ranging from $\sim$5 $\times$ 10$^{6}$ cm$^{-3}$ 
(SM2) to $\sim$4 $\times$ 10$^{7}$ cm$^{-3}$ (SM1N), i.e., in the regime 
where central N$_{2}$H$^{+}$ depletions were found by AOH03.  Note, however, 
that the AOH03 models had constant $T_{K}$ = 10 K, so the possible effect 
on depletion that warmer $T_{dust}$ may have must also be considered in an 
intepretation. 

Table 4 lists $X$(N$_{2}$H$^{+}$), $N$(N$_{2}$H$^{+}$), and $N$(H$_{2}$) 
for Oph A-N6 in addition to those for VLA 1623, SM1N, SM1, and SM2, also
found using Equations 2 and 3 and averaging over coincident beam-sized pixels.  
For N6, we assume $T_{dust}$ = 15 K following AWB93, although they suggested 
that $T_{dust}$ $\leq$ 15 K southeast of SM2.  The $N$(N$_{2}$H$^{+}$) of N6, 
6 $\times$ 10$^{13}$ cm$^{-2}$, lies between the higher values of SM1 and SM1N 
and the lower values of SM2 and VLA 1623.  On the other hand, the $N$(H$_{2}$) 
of N6, $\sim$3 $\times$ 10$^{23}$ cm$^{-2}$, is actually similar to those of
SM1N and SM1, but appears smaller than those of VLA 1623 and SM2.  

Relative to Oph A-N6, VLA 1623 and SM2 have likely higher mean densities,
given their apparently larger $N$(H$_{2}$) and more compact appearance as 
1.3 mm continuum maxima.  In the context of the AOH03 models, their low
N$_{2}$H$^{+}$ column densities are due to N$_{2}$H$^{+}$ depletion at high
density.  Since VLA 1623 has formed a protostar (or two; see Looney, Mundy, 
\& Welch 2000) but SM2 not formed any, the mean density of VLA 1623 should 
be even larger than that of SM2.  The weaker N$_{2}$H$^{+}$ emission of VLA 
1623 compared to SM2 likely reflects even greater N$_{2}$H$^{+}$ depletion 
that should accompany higher density.  (In addition, N$_{2}$H$^{+}$ depletion
toward VLA 1623 may be enhanced if grain mantles containing CO have returned 
to the gas phase through evaporation or outflow shocks.)  SM1N and SM1 do not 
fit easily into this sequence, however, as they have higher $N$(N$_{2}$H$^{+}$) 
but similar $N$(H$_{2}$).  Their higher $N$(N$_{2}$H$^{+}$) may reflect their
warmer $T_{dust}$ relative to SM2 if N$_{2}$ does not easily adsorb onto 
their warmer dust grains.  CO, however, is likely also not adsorbed readily 
at these $T_{dust}$ values, and its presence in the gas phase likely would 
reduce $N$(N$_{2}$H$^{+}$) rather than increase it.  The relatively high 
$N$(N$_{2}$H$^{+}$) for SM1N and SM1 may be due to internal, dense regions 
where CO has been depleted but not N$_{2}$, i.e., that are cooler than 
suggested by their (average) $T_{dust}$ but warmer than regions of similar
density in SM2.

Since density must increase as a result of evolution towards star formation
and $X$(N$_{2}$H$^{+}$) appears to decrease above $\sim$10$^{6}$ cm$^{-3}$, 
$X$(N$_{2}$H$^{+}$) may be useful as an indicator of the relative evolution 
of several objects forming within the same dense core.  Our $X$(N$_{2}$H$^{+}$) 
results alone therefore suggest that Oph A-N6 is at an earlier stage relative 
to SM2 and that SM2 is at an earlier stage relative to VLA 1623.  Given the 
warmer $T_{dust}$ of SM1N and SM1 relative to SM2, their evolutionary stage
relative to N6, SM2, and VLA 1623 is difficult to judge.  We stress that these 
results are highly qualitative.  Evans et al. (2001) and Zucconi, Walmsley, 
\& Galli (2001) have shown recently that Bonnor-Ebert spheres with radial 
temperature gradients (e.g., with low central $T_{dust}$) are better than 
isothermal models in reproducing the observed dust emission of isolated 
starless objects, and the $T_{dust}$ values derived by AWB93 and used here 
are likely averages of the actual dust temperature distributions within the 
Oph A objects.  More detailed modeling of the dust continuum emission of 
all these objects is clearly required to determine better their thermal and 
density structures and provide more quantitative results, especially if 
N$_{2}$ depletion onto dust grains is strongly dependent on $T_{dust}$.

\subsection{Nonthermal Motions and Velocity Gradients}

Variations of the $\Delta V$ and $V_{LSR}$ of N$_{2}$H$^{+}$ 1--0 in Oph A 
allow us to probe the dynamics of dense gas within the core.  Complementing 
the previous discussion on N$_{2}$H$^{+}$ abundances, the dynamics of dense
gas provide further insight into the relative evolution of objects within 
Oph A.  In the following, we consider separately the dynamics of Oph A as a 
whole and of its population of starless objects, as well as those of 
Oph A-N6 relative to SM1N, SM1, and SM2.

\subsubsection{The Oph A Core} 

The effective average velocity dispersion of gas, $\sigma$, within Oph A 
can be estimated from our data and compared to simple models to evaluate the 
dynamical state of the core.  Following Bertoldi \& McKee (1992; Appendix C), 
$\sigma$ can be estimated in several ways.  First, the dispersion of an 
average spectrum comprised of all line data across the core can be measured 
and $\sigma^{2}$ is approximately equal to the quadrature sum of this quantity 
and the thermal dispersion, i.e., that of velocities of mean molecular mass 
expected from thermal motions.  The FWHM resulting from a Gaussian fit to 
such an average spectrum from our Oph A data is 0.58 km s$^{-1}$, yielding 
$\sigma$ = 0.36 km s$^{-1}$ if $T_{K}$ = 20 K.  Second, the rms of both 
$\Delta V$ and $V_{LSR}$ can be measured and $\sigma^{2}$ is approximately 
equal to the sum of the squares of these quantities and the thermal 
dispersion.  Using values from Table 1 and again assuming $T_{K}$ = 20 K, 
$\sigma$ = 0.33 km s$^{-1}$, i.e., within 10\% of the first value.

The estimated $\sigma$ of Oph A is similar to but less than expected from simple 
virial arguments, i.e., those that ignore contributions from surface pressure 
and magnetic fields\footnote{Significant surface pressure would increase this 
dispersion but significant magnetic field energy would decrease it.  Within 
Ophiuchus, however, magnetic field strengths appear small; e.g., $B_{\|}$ 
$\leq$ 10 $\mu$G was found in Oph from OH Zeeman measurements by Troland et 
al. (1996).}.  For example, the average total one-dimensional velocity 
dispersion of a virialized core can be estimated by equating 2$T$ = 2(3/2)$M 
\sigma^{2}$, the magnitude of the total core kinetic energy ($M$ is the 
total core mass), and $W$ = $(3/5R)aGM^{2}$, the magnitude of the total core 
gravitational potential energy ($R_{m}$ is the average projected core size, 
$G$ is the gravitational constant, and $a$ is a dimensionless parameter 
accounting for core asphericity and nonuniformity; see Bertoldi \& McKee).  
For Oph A, $M$ $\approx$ 11 M\sun, assuming $T_{dust}$ = 20 K on average 
and using the 1.3 mm continuum data of MAN98 over positions with detected 
N$_{2}$H$^{+}$ 1--0 emission.  We find $\sigma$ = 0.57 km s$^{-1}$, assuming 
this mass and uniformity in density.  (An aspect ratio of $\sim$7 and a 
projected semiminor axis of $\sim$0.01 pc are also assumed, yielding $R_{m}$ 
= 0.023 pc and $a$ = 0.78; see Fig. 2 of Bertoldi \& McKee.)  These velocities 
suggest Oph A as a whole is in gravitational virial equilibrium to first order, 
i.e., they are within a factor of 2 of those required.  (Although these 
velocities suggest Oph A has insufficient support against gravity in the 
absence of significant magnetic energy, the difference is likely not 
significant given the uncertainties in our analysis.)  AWB93 also concluded 
the Oph A filament was similarly close to gravitational virial equilibrium, 
based on their submillimeter continuum data and a typical $\Delta V$ from 
available line data.  More recently, Tachihara et al. found similar conclusions 
for (lower density) star forming C$^{18}$O cores in the near-neighborhood of 
Oph A.  Better determinations of the local magnetic field strengths are clearly 
needed to evaluate better the global dynamical state of Oph A.

\subsubsection{The Starless Object Population}

Variations of $V_{LSR}$ in the population of objects within Oph A provide 
constraints on some models of clustered star formation.  Table 4 lists 
the $V_{LSR}$ of SM1N, SM1, SM2, and Oph A-N6, obtained by averaging 4-5 
beam-sized pixels around each object.  The rms of these values is only 0.10 
km s$^{-1}$, or $\sim$0.4 $c_{s}$(20 K), the isothermal sound speed of 
molecular gas at $T_{gas}$ = 20 K.  The similarities between the $V_{LSR}$ 
of gas associated with the continuum objects and Oph A-N6 suggest strongly 
that all these lie within the same filament, and are not objects in separate 
cores seen by coincidence near each other towards Ophiuchus.  Furthermore, 
the objects do not have substantial radial velocities relative to their 
neighboring dense gas.  For example, the mean difference in the $V_{LSR}$ 
of a given Oph A object and the mean of its surrounding gas is only 0.06 km 
s$^{-1}$, much smaller than the rms of $V_{LSR}$.  This value is only 
$\sim$0.25 $c_{s}$(20 K), i.e., at the low end of expected mean relative 
gas velocities in cluster formation models where starless objects move 
ballistically within a core and accrete mass competitively (e.g., see 
Bonnell et al. 1997, 2001).  The Oph A objects, however, are few in number, 
are found in close proximity, and as a whole may be at a relatively early, 
slow stage of such an evolution (e.g., when gas dominates the gravitational 
potential). 

\subsubsection{Oph A-N6}

Variations of $\Delta V$ between starless objects in Oph A may be also 
related to these objects being at different stages in the evolution of 
dense gas towards star formation.  Here we consider Oph A-N6 as an object,
to place its physical characteristics in the context of models relevant to 
these epochs.  Unlike SM1, SM1N, and SM2, N6 is not a local 2D maximum of 
continuum emission (i.e., $N$(H$_{2}$)), but instead resides on a gradient 
of continuum emission along the local northwest-southeast direction of the 
filamentary N$_{2}$H$^{+}$ 1--0 emission.  Note, however, that N6 is at least 
at a 1D maximum of continuum emission in the direction perpendicular to the 
filament.

Given its 125 pc distance and following our arbitrary definition of its 
extent, i.e., where $\Delta V$ $\leq$ 0.30 km s$^{-1}$, Oph A-N6 has a 
(nondeconvolved) geometrical mean radius of 0.008 pc.  From its 1.3 mm 
continuum emission (see Fig. 1a) and assuming its dust is isothermal at 
$T_{dust}$ = 15 K, N6 has a mean column density of 2.5 $\times$ 10$^{23}$ 
cm$^{-2}$ and a mass of 0.9 M\sun.  (Assuming N6 is a sphere, its mean 
density is 7.6 $\times$ 10$^{6}$ cm$^{-2}$, well above the 2.5 $\times$ 
10$^{5}$ critical density of the N$_{2}$H$^{+}$ 1--0 line.)  In addition, 
N6 has a mean $\Delta V$ = 0.25 km s$^{-1}$ (see Table 4), suggesting a 
mean dispersion for the N$_{2}$H$^{+}$ 1--0 line of 0.11 km s$^{-1}$.  
Assuming the gas of N6 is isothermal at $T_{gas}$ = 17.4 K (see \S 3), 
the mean nonthermal component $\sigma_{NT}$ = 0.080 km s$^{-1}$ and 
$\sigma$ = 0.26 km s$^{-1}$. 

The observed quantities of Oph A-N6 are within a factor of 2 of those
predicted from simple models of equilibrium spheres.  For example, the
mass-radius ratio of a uniform sphere in gravitational virial equilibrium
is $5\sigma^{2}/G$ and that of a critically-stable Bonnor-Ebert sphere is 
$2.4\sigma^{2}/G$ (Bonnor 1956).  Given the effective radius and mass of 
N6, the mean velocity dispersions predicted from these simple models are 
0.31 km s$^{-1}$ and 0.45 km s$^{-1}$ respectively, i.e., more than that of 
N6 itself by factors of 1.2 and 1.7 respectively.  Like Oph A as a whole, 
N6 is in gravitational virial equilibrium to first order.  (Again, the 
differences are likely not significant enough to suggest N6 has insufficent 
support against gravity.)  Interestingly, the mass and radius of N6 can be 
reproduced simultaneously if the mass-radius ratio $\approx$ $7\sigma^{2}/G$, 
a situation that could occur, e.g., in virial equilibrium if magnetic 
energy $B$ $\approx$ $(6/5)M\sigma^{2}$, or more if surface pressure is 
also non-zero.  

There are three questions about Oph A-N6 that simple equilibrium models do 
not directly address.  First, why does N6 have such narrow line widths that 
suggest it has relatively little turbulence?  Second, why are these line 
widths narrower than those in its immediate surroundings?  Third, why is 
N6 located so close to its neighbor SM2?  These questions may be answered 
simultaneously by considering models of equilibrium spheres embedded in 
sufficiently turbulent and opaque cores.  Within such environments, turbulent 
motions could be reduced locally, e.g., if the spheres were similar or smaller 
in size than the cutoff wavelength for MHD waves in their surroundings (McKee 
1989, McKee et al. 1993).  (Turbulence could instead be damped locally by other 
processes such as shocks or dissipation; e.g., see Nakano 1998.)  Myers (1999; 
see also Myers 1997), following reasoning by Larson (1985) and Mouschovias 
(1991), proposed thermally dominated, critically stable Bonnor-Ebert spheres 
could exist within turbulent, opaque, and dense cores.  Such cores require 
large column densities (e.g., $A_{V}$ $>$ 4), so that ionization by cosmic 
rays dominates over that by ultraviolet radiation, and the coupling between 
the magnetic field and the gas is minimized.  The embedded objects, called 
``kernels" by Myers, would be compressed by external (turbulent) core pressure 
to sizes comparable to the observed separations of young stellar objects within 
embedded clusters (e.g., $\sim$0.05-0.15 pc).  A protostellar cluster could 
then arise from the near-simultaneous collapse of several kernels within a 
common core.

Oph A-N6 may be an excellent candidate for a kernel-like object, given its
near-thermal line widths.  Furthermore, its 0.9 M\sun\ mass matches quite 
well the 0.8 M\sun\ mean mass predicted for kernels within nearby star forming 
molecular cores, i.e., masses obtained by Myers using values of $T$, $\Delta 
V_{NT}$ and $n$ derived by from Harju, Walmsley, \& Wouterloot (1993) and 
Ladd, Myers, \& Goodman (1994).  Also, the projected separation between N6 and 
SM2 is 0.025 pc, within a factor of $\sim$2 of the $\sim$0.05 pc separation 
of kernels predicted by Myers.  (Indeed, the mean nearest-neighbor separation 
of N6, SM1N, SM1, SM2, and VLA 1623, is only 0.023 pc.)  Oph A-N6 does not 
fit completely to a kernel interpretation, however.  As noted above, its 
0.008 pc radius is $\sim$3 smaller than the 0.024 pc mean radius expected 
for such kernels.  (Note that the radius of a singular isothermal sphere (Shu 
1977) at $T$ = 17.4 K required to make a 0.9 M\sun\ star is $\sim$0.032 pc, 
a factor of $\sim$4 larger than that of N6.)  Despite these caveats, however, 
the kernel model provides at least some basis for understanding N6 as a 
highly localized region of low turbulence, unlike simple equilibrium models.

The non-detection of Oph A-N6 as a two-dimensional maximum of $N$(H$_{2}$)
but its clear detection as a location of maximal $N$(N$_{2}$H$^{+}$) suggest 
that interesting sites possibly related to star formation may be overlooked 
in maps of thermal continuum emission.  Why was N6 not seen, e.g., by MAN98?  
Contrasts of $N$(H$_{2}$) in N6 could be minimized if dust at its inner radii 
are colder than dust at its outer radii (e.g., see Evans et al.), a strong 
possibility given the high densities and extinctions observed in Oph A.  Note, 
however, that Bouwman, Galli, \& Andr\'e (2004, in preparation; see also Andr\'e 
et al.  2003) have recently shown that 1.3 mm continuum flux and $T_{dust}$ 
are anticorrelated in starless cores, even those with the high column densities 
and incident radiation fields of Oph A (e.g., 85 $G_{o}$; Liseau et al. 1999).  
The Oph A objects, especially N6, must be specifically modeled to evaluate 
better this possibility.

%
%

If Oph A-N6 is a kernel-like object, we speculate that the Oph A continuum 
objects may also be or once were kernel-like.  In this scenario, the large 
$\Delta V$ of these objects relative to that of Oph A-N6 may suggest that 
they are now in a state of collapse.  For example, N6 has narrow lines 
close to the thermal line width; an absence of wings or asymmetrically blue 
profiles in the line suggest that collapse has not yet occurred.\footnote{If 
N$_{2}$H$^{+}$ 1--0 has thermalized within N6, and $T_{ex}$ $\approx$ $T_{K}$, 
the line will not show infall asymmetry if no gradient in $T_{ex}$ exists.}  
If the continuum objects have had similar physical conditions to N6 earlier, 
turbulence may have been similarly reduced at that epoch.  Therefore, their 
larger $\Delta V$ may not be due to substantial turbulent motions in their 
associated gas.  Indeed, how turbulent motions could be reintroduced to these 
objects after its reduction in the absence of protostars, winds, or outflows 
is unclear.  Instead, gas within these objects may still be thermally dominated 
and the line broadening observed may be due to infall motions stemming from 
collapse\footnote{Belloche et al. (2001; 2004, in preparation) have detected 
the asymmetrically blue profiles of CS lines that suggest infall toward at 
least SM2.}.  The N$_{2}$H$^{+}$ 1--0 line is only marginally optically thick 
toward these objects, so the line should remain symmetric, as seen in Figure 5, 
if emitted from infalling gas (Myers, Evans, \& Ohashi 2000).  We note here 
again that SM1, SM1N, and SM2 are found relatively close to (but are not 
coincident with) positions of the largest $\Delta V$ in Oph A; these extreme 
line widths may also be caused by infall motions related to these objects. 
Furthermore, the correlation of $\Delta V$ vs.  $V_{LSR}$ found for bright 
line emission shown in Fig. 6a may be also related to infall motions.  The 
observed FWZIs of N$_{2}$H$^{+}$ 1--0 toward SM1N, SM1, and SM2 are $\sim$1.0 
km s$^{-1}$, suggesting infall motions on scales of $\sim$1000-3000 AU up to 
0.5 km s$^{-1}$, similar to those measured directly toward the Class 0 objects 
NGC 1333 IRAS 4A and 4B by Di Francesco et al. 

Further speculating, we note that the variations of $V_{LSR}$ in Oph A may 
provide additional evidence for differences in the evolutionary stages of 
starless objects within Oph A, by supporting the idea that Oph A-N6 is not 
collapsing but SM1N, SM1, and SM2 may be.  N6 appears to have no associated 
N$_{2}$H$^{+}$ 1--0 velocity gradients but the continuum objects have small 
ones (i.e., $\leq$0.5 km s$^{-1}$ over $\sim$0.01 pc; see Fig. 3b).  Such 
small velocity gradients could stem from increases in local rotation speeds 
of dense gas that could accompany inward motions due to angular momentum 
conservation.  (Indeed line broadening by rotational motions may also be 
contributing to the larger line widths associated with these objects.  Note 
that the gradients in Fig. 3b, however, are seen on scales larger than the 
beam size.)  High-resolution data of Oph A in other tracers of quiescent 
dense gas that require higher levels of excitation than N$_{2}$H$^{+}$ 1--0 
could reveal velocity gradients more dominated by infall or rotation and 
clarify this issue (e.g., see Caselli et al. 2002b).

\section{Summary and Conclusions}

1. Combined single-dish and interferometer data of N$_{2}$H$^{+}$ 1--0 
line emission from Oph A are presented that yield a map of dense gas at 
high linear resolution (e.g., $\sim$1000 AU) with information retained 
on large spatial scales.  Six maxima of integrated line intensity, Oph 
A-N1 through N6, have been detected.  N4 and N5 are coincident with the 
starless objects SM1 and SM1N respectively.  The other maxima, however, 
are not coincident with known starless or protostellar Oph A objects.  
Moreover, little N$_{2}$H$^{+}$ 1--0 emission is associated with the 
starless object SM2 or the Class 0 protostar VLA 1623.  This situation
is in contrast with observations of isolated cores where dust peaks and 
integrated N$_{2}$H$^{+}$ intensity maxima are commonly coincident.

2.  With 1.3 mm continuum data from MAN98 and $T_{dust}$ estimates from
AWB93, $N$(N$_{2}$H$^{+}$), $N$(H$_{2}$), and $X$(N$_{2}$H$^{+}$) were 
estimated for the Oph A objects.  SM1 and SM1N have similar values of
$X$(N$_{2}$H$^{+}$) $\approx$ 2-3 $\times$ 10$^{-10}$.  SM2 and VLA 1623, 
however, have values of $X$(N$_{2}$H$^{+}$) that are factors of $>$4 less 
than those of SM1 and SM1N.  The $X$(N$_{2}$H$^{+}$) of Oph A-N6 is similar 
to that of SM1 and SM1N, if $T_{dust}$ $\approx$ 15 K at that location.  
For SM2 and VLA 1623, the lower values of $X$(N$_{2}$H$^{+}$) are likely 
due to gas-grain chemistry in cold, dense cores where N$_{2}$ adsorbs onto 
grain surfaces and N$_{2}$H$^{+}$ is depleted, as suggested by many.  
Similar depletion may not have occurred at SM1 or SM1N because their dust 
is warmer, e.g., $T_{dust}$ = 27 K, perhaps inhibiting N$_{2}$ adsoption.

3. N$_{2}$H$^{+}$ 1--0 line widths ($\Delta V$) in Oph A vary by a factor 
of $\sim$5 with values $<$0.30 km s$^{-1}$ found in fourteen locations.
Of these, only the location associated with the N6 maximum of integrated 
intensity is both a local minimum of line width and has an extent larger 
than the beam size.  The minimum $\Delta V$ in N6 is only 0.193 km s$^{-1}$, 
or just larger than line widths expected from thermal motions alone at 
expected temperatures.  In contrast, the $\Delta V$ associated with the 
SM1, SM1N, and SM2 objects are $\sim$0.5 km s$^{-1}$.  

4.  N$_{2}$H$^{+}$ 1--0 line centroids ($V_{LSR}$) in Oph A vary little 
across the core, having an rms of only $\sim$0.17 km s$^{-1}$.  No 
large-scale gradient is seen across Oph A, although small-scale $V_{LSR}$ 
gradients of $<$0.5 km s$^{-1}$ over 0.01 pc are associated with SM1, SM1N, 
and SM2, but not with N6.  The $\Delta V$ and $V_{LSR}$ of Oph A as a whole 
yield a one-dimensional velocity dispersion, $\sigma$, for Oph A that is a 
factor of $\sim$2 of that required for gravitational virial equilibrium.  

5.  The low $\Delta V$ and lack of $V_{LSR}$ gradient of N6 relative to 
SM1, SM1N, and SM2 suggest evolutionary differences.  The one-dimensional 
velocity of N6 suggest it alone is within a factor of 2 from gravitational 
virial equilibrium.  N6 has properties (e.g., $\Delta V$, mass, and short
proximity to neighbors) similar to those predicted for ``kernels", i.e., 
thermally dominated, critically stable, Bonnor-Ebert spheres embedded within 
turbulent cores, although some of its properties (e.g., size and no column 
density maximum) are dissimilar.  N6 may represent an early stage of the 
evolution of dense gas towards star formation where significant local 
reduction of turbulent motions occurs.  The non-detection of N6 in 
thermal continuum maps of Oph A suggests that interesting sites possibly 
related to star formation may be missed in such maps.


\acknowledgements{We thank our referee, Edwin Bergin, for numerous suggestions
which improved this paper.  In addition, we thank David Wilner, Frederique Motte, 
Aurore Bacmann, Tamara Helfer, Zhi-Yun Li, Fumitaka Nakamura, and Helmut Weisemeyer 
for fruitful discussions regarding this project.  Furthermore, JD thanks Leo Blitz 
and the Radio Astronomy Laboratory of the University of California, Berkeley for 
generous support from a postdoctoral fellowship via the Berkeley-Illinois-Maryland 
Association.}

\clearpage

\figcaption{a) 1.3 mm continuum emission from Oph A mapped by MAN98 with the 
IRAM 30 m Telescope.  Solid contours start at 100 mJy beam$^{-1}$, increase 
in steps of 100 mJy beam$^{-1}$ to 1400 mJy beam$^{-1}$.  The greyscale range 
shown is 75--2000 mJy beam$^{-1}$.  The 1 $\sigma$ rms noise level of the map 
is $\sim$10 mJy beam$^{-1}$.  Significant emission is found at lower levels 
than indicated by the first contour, but this has been cut here for clarity.  
The position of the Class 0 protostar VLA 1623 from MAN98 is denoted by a box, 
and positions of starless objects from MAN98 are denoted by triangles.  The 
position of the near by B star S1 (also E25; Elias 1978) is denoted by a star.  
The black circle at lower right denotes the relative size of the 13\as\ map 
resolution.  b) Integrated intensity of N$_{2}$H$^{+}$ 1--0 line emission from 
Oph A obtained with the IRAM 30 m Telescope and BIMA interferometer.  The 
values shown are the ``zeroth-moment" of CLEANed data, where only intensities 
exceeding $\pm$2 $\sigma$ rms per channel were summed.  Solid contours denote 
2.8, 8.4, 14.0, 19.7, 25.3, 30.9, 36.5, 42.1, and 47.8 K km s$^{-1}$, while 
the dashed contour denotes -2.8 K km s$^{-1}$.  The greyscale range shown 
is 1.4--65.2 K km s$^{-1}$.  The 1 $\sigma$ rms noise level over the central 
mosaic area is 1.4 K km s$^{-1}$.  Symbols are defined as for panel a), except 
the position of maximum N$_{2}$H$^{+}$ integrated intensity Oph A-N6 is denoted 
as an open diamond.  Also, the ellipse at lower right here denotes the relative 
size of the 9\farcs9 $\times$ 6\farcs1 map resolution.  
\label{fig1}}

\figcaption{a) Column densities of H$_{2}$ obtained using Equation 2 and the 
1.3 mm continuum data of MAN98 assuming a constant $T_{dust}$ = 20 K at all 
locations.  The values shown are those derived after binning the MAN98 data 
first into 13\farcs5 $\times$ 13\farcs5 pixels (i.e., approximately the beam 
size of the MAN98 1.3 mm continnum IRAM 30 m data) and then regridding into 
6\farcs75 $\times$ 10\farcs5 pixels (i.e., approximately the beam size of our 
N$_{2}$H$^{+}$ line BIMA+IRAM data.)  The $N$(H$_{2}$) shown have been divided 
by 10$^{21}$ cm$^{-2}$, and the color scale ranges from 0 to 531.  Triangles 
denote the locations of starless continuum objects detected by MAN98, the box 
denotes the location of VLA 1623, and the diamond denotes the location of Oph
A-N6.  The white contour displays locations in the N$_{2}$H$^{+}$ line data 
where $T_{mb}^{max}$ $\geq$ 3 $\sigma$.
b) Column densities of N$_{2}$H$^{+}$ obtained using Equation 3 and values 
of $\tau_{tot}$, $T_{ex}$, and $\Delta V$ obtained from HFS fits.  The values 
shown are those derived after binning the combined data from 0\farcs5 $\times$ 
0\farcs5 pixels into 6\farcs75 $\times$ 10\farcs5 pixels (i.e., approximately 
the beam size of the data.)  The $N$(N$_{2}$H$^{+}$) shown have been divided 
by 10$^{12}$ cm$^{-2}$ and the color scale ranges from -12 to 125.  Symbols and 
contours are defined as for Figure 3a.
c) Fractional abundances of N$_{2}$H$^{+}$ obtained by assuming $N$(H$_{2}$) 
is linearly proportional to continuum dust emission and $T_{dust}$ = 20 K for 
every pixel (see Equation 2).  The $X$(N$_{2}$H$^{+}$) shown have been divided 
by 10$^{11}$ and the color scale ranges from 0 to 64.  Symbols and contours 
are defined as for Figure 3a.  \label{fig2}}

\figcaption{a) Spatial distribution of the characteristics of N$_{2}$H$^{+}$ 
1--0 from Oph A measured by HFS fitting the combined IRAM and BIMA data.  Only 
results from data where the peak intensity of the brightest hyperfine component 
equalled or exceeded 3 $\sigma$ rms are shown.  Symbols are defined as for 
Figure 1.  a) Line width (FWHM) or $\Delta V$.  The color scale shown ranges 
from 0.2 to 0.8 km s$^{-1}$ FWHM.  The contours shown begin at 0.15 km s$^{-1}$
and increase in steps of 0.15 km s$^{-1}$ to 0.90 km s$^{-1}$.  b) Line velocity 
or V$_{LSR}$.  The color scale range shown is 3--4 km s$^{-1}$.  The contours 
shown begin at 2.8 km s$^{-1}$ and increase in steps of 0.2 km s$^{-1}$ to 4.2 
km s$^{-1}$.  \label{fig3}} 

\figcaption{The observed spectrum of N$_{2}$H$^{+}$ 1--0 at the position of 
minimum line width in Oph A-N6, i.e., at R.A., decl. (2000) = 
16$^{h}$26$^{m}$30.75$^{s}$, -24\deg 24\am 40\farcs45 (solid line), together 
with a spectrum corresponding to its HFS fit (dashed line). \label{fig4}}

\figcaption{Spectral grid of N$_{2}$H$^{+}$ 101-012 (i.e,. the ``isolated" 
component of N$_{2}$H$^{+}$ 1--0) over Oph A.  Each panel contains a spectrum 
corresponding to the combined N$_{2}$H$^{+}$ data averaged into pixels 
10\farcs5 $\times$ 6\farcs75 in size (i.e., approximately the size of the 
combined beam.)  The panels are positioned to correspond to previous images 
with this pixel sizes, e.g., Figure 3.  The abscissa of each panel is 
$V_{LSR}$ and ranges from 2.0 km s$^{-1}$ to 6.0 km s$^{-1}$.  The ordinate 
of each panel is $T_{B}$ and ranges from -3 K to 15 K.  Each spectral 
channel is 0.079 km s$^{-1}$ wide.  Triangles, squares, and diamonds are 
defined as in previous Figures. \label{fig5}}

\figcaption{a) Plot of $V_{LSR}$ vs. $\Delta_V$ for N$_{2}$H$^{+}$ 1--0 in 
Oph A from HFS fits to combined data in beam-sized pixels, where T$^{B}_{max}$ 
$>$ 9 $\sigma$.  The solid line is the linear fit to these data.  b) the same 
quantities as shown in Figure 9a, except the data shown are those from where 
T$^{B}_{max}$ is 3--9 $\sigma$.  \label{fig6}}

\figcaption{Spectra of N$_{2}$H$^{+}$ 101-012 averaged over various locations 
in Oph A.  {\it Bold-line}\/ spectra are averages over 2-4 neighboring 
beam-sized pixels containing the narrowest bright N$_{2}$H$^{+}$ emission and 
the {\it thin-line}\/ spectra are averages over the brightest 4-6 surrounding 
pixels with weaker, broader emission.  Spectra 1 consist of narrow-line 
emission averaged over 2 pixels in Oph A-N6 or broader-line emission averaged 
over 4 pixels around Oph A-N6.  Spectra 2 consist of narrow-line emission 
averaged over 3 pixels north of VLA 1623 or east of SM1 or broader-line 
emission averaged over 7 pixels to the north, west, or south of the 
narrow-line emission.  Spectra 3 consist of narrow-line emission in 4 
pixels northwest of SM1N or broader-line emission in 2 pairs of 2 pixels 
northwest or southeast of the narrow-line emission.  \label{fig7}}

\clearpage

\begin{table*}
\caption{Positions of Maximum N$_{2}$H$^{+}$ 1--0 Integrated Intensity in Oph A}
\label{tbl-1}

\begin{center}
\begin{tabular}{cccc}
\tableline
Name & R.A.(2000) & decl. (2000) & Association  \cr
\tableline
\tableline
N1   & 16 26 26.4 & -24 23 07    &  A-MM5?      \cr
N2   & 16 26 26.5 & -24 22 24    &  \nodata     \cr
N3   & 16 26 27.2 & -24 24 20    &  \nodata     \cr
N4   & 16 26 27.2 & -24 23 50    &  SM1         \cr
N5   & 16 26 27.4 & -24 23 30    &  SM1N        \cr
N6   & 16 26 31.6 & -24 24 52    &  \nodata     \cr
\tableline
\end{tabular}
\end{center}
\end{table*}


\begin{table*}
\caption{N$_{2}$H$^{+}$ Line Characteristics in Oph A}
\label{tbl-2}

\begin{center}
\begin{tabular}{cccccc}
\tableline
Value         & Unit         & mean  & rms   & min       & max  \cr
\tableline
\tableline
$V_{LSR}$     & km s$^{-1}$  &  3.49 &  0.17 &    2.98   &  3.91  \cr

$\Delta V$    & km s$^{-1}$  &  0.39 &  0.11 &    0.14   &  0.88  \cr

$T_{B}$       & K            & 13.68 &  3.47 &    5.25   & 26.29  \cr

$\tau_{tot}$  &              &  5.40 &  3.33 & $\leq$0.1 & 13.05  \cr
\tableline
\end{tabular}
\end{center}
\end{table*}


\begin{table*}
\caption{Column Densities and N$_{2}$H$^{+}$ Abundances in Oph A}
\label{tbl-3}

\begin{center}
\begin{tabular}{cccccc}
\tableline
Value               & Unit                & mean  & rms   & min   & max  \cr
\tableline
\tableline
$N$(H$_{2}$)        & 10$^{21}$ cm$^{-2}$ & 179.7 & 125.3 & 32.44 & 530.8 \cr

$N$(N$_{2}$H$^{+}$) & 10$^{12}$ cm$^{-2}$ & 30.24 & 27.83 & 2.675 & 125.9 \cr

$X$(N$_{2}$H$^{+}$) & 10$^{-11}$          & 18.61 & 12.73 & 1.533 & 64.34 \cr
\tableline
\end{tabular}
\end{center}
\end{table*}

\begin{table*}
\caption{N$_{2}$H$^{+}$ Characteristics of Oph A Objects}
\label{tbl-4}

\begin{center}
\begin{tabular}{ccccccc}
\tableline
Object   & $T_{dust}$(AWB93)& $X$(N$_{2}$H$^{+}$) & $N$(N$_{2}$H$^{+}$) & $N$(H$_{2}$)\tablenotemark{1} & $\Delta V$(N$_{2}$H$^{+}$) & $V_{LSR}$ \cr
& K & $\times$10$^{-10}$ & $\times$10$^{13}$ cm$^{-2}$ & $\times$10$^{23}$ cm$^{-2}$ & km s$^{-1}$ & km s$^{-1}$ 
\cr
\tableline
\tableline
VLA 1623 & 15-20 & $<$0.4--$<$0.7 & $<$3  & 6-4  & \nodata & \nodata \cr
SM1N     & 27    & 3              & 10    & 3    & 0.49    & 3.60   \cr
SM1      & 27    & 2              & 8     & 4    & 0.59    & 3.72   \cr
SM2      & $<$20 & $<$0.9         & 2     & $>$3 & 0.52    & 3.47   \cr
Oph A-N6 & 15?   & 4              & 6     & 3    & 0.25    & 3.47   \cr

\tableline
\tablenotetext{1}{$N$(H$_{2}$) values were derived assuming all continuum
emission detected along the lines-of-sight are associated with the objects 
in question.  Since Oph A-N6 is not an peak of continuum emission like the 
other objects here, its $N$(H$_{2}$) may be overestimated.}

\end{tabular}
\end{center}
\end{table*}


\begin{table*}
\caption{Positions in Oph A of N$_{2}$H$^{+}$ 1--0 $\Delta V$ $<$ 0.30 km s$^{-1}$}
\label{tbl-5}

\begin{center}
\begin{tabular}{ccccc}
\tableline
Name & R.A.(2000) & decl. (2000) & $\Delta V$  &  Association  \cr
\tableline
\tableline
a    & 16 26 24.8 & -24 22 54    &   0.24   &  \nodata      \cr
b    & 16 26 26.1 & -24 23 14    &   0.29   &    N1         \cr
c    & 16 26 26.3 & -24 22 24    &   0.26   &    N2         \cr
d    & 16 26 26.3 & -24 24 08    &   0.27   &  \nodata      \cr
e    & 16 26 26.4 & -24 23 48    &   0.28   &  \nodata      \cr
f    & 16 26 26.9 & -24 23 10    &   0.29   &  \nodata      \cr
g    & 16 26 29.3 & -24 24 40    &   0.27   &    N1         \cr
h    & 16 26 30.0 & -24 25 01    &   0.28   &  \nodata      \cr
i    & 16 26 30.1 & -24 24 05    &   0.28   &  \nodata      \cr
j    & 16 26 30.6 & -24 25 01    &   0.29   &  \nodata      \cr
k    & 16 26 30.7 & -24 24 40    &   0.27   &  \nodata      \cr
l    & 16 26 30.7 & -24 25 20    &   0.26   &    N6         \cr
m    & 16 26 32.4 & -24 24 33    &   0.21   &  \nodata      \cr
n    & 16 26 33.4 & -24 24 48    &   0.29   &  \nodata      \cr
\tableline
\end{tabular}
\end{center}
\end{table*}

\clearpage

\begin{figure}
\vspace{7.25in}
\hspace{-0.25in}
\includegraphics{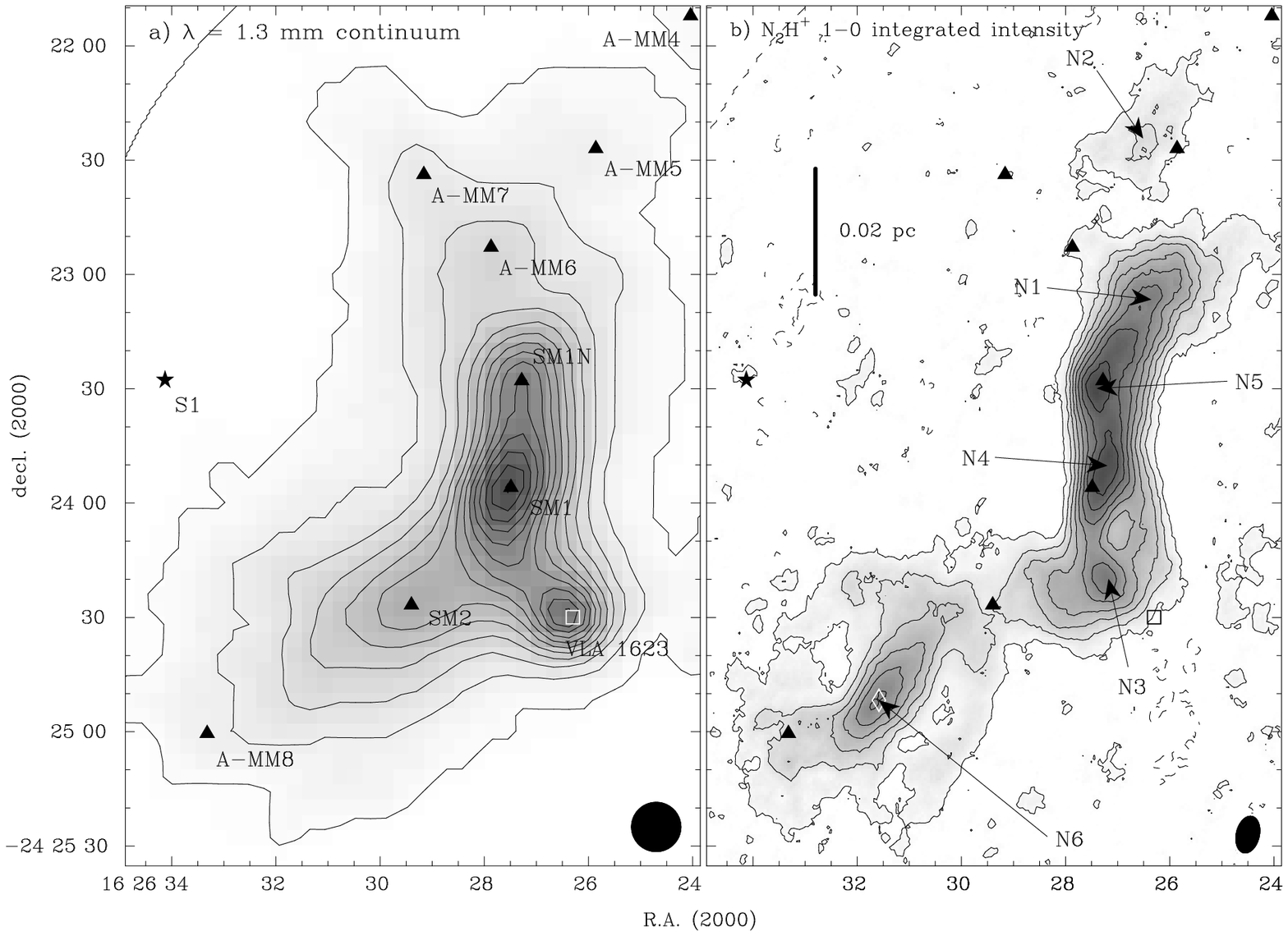}
\vspace{-2.25in}
\end{figure}

\begin{figure}
\vspace{7.25in}
\hspace{-0.25in}
\includegraphics{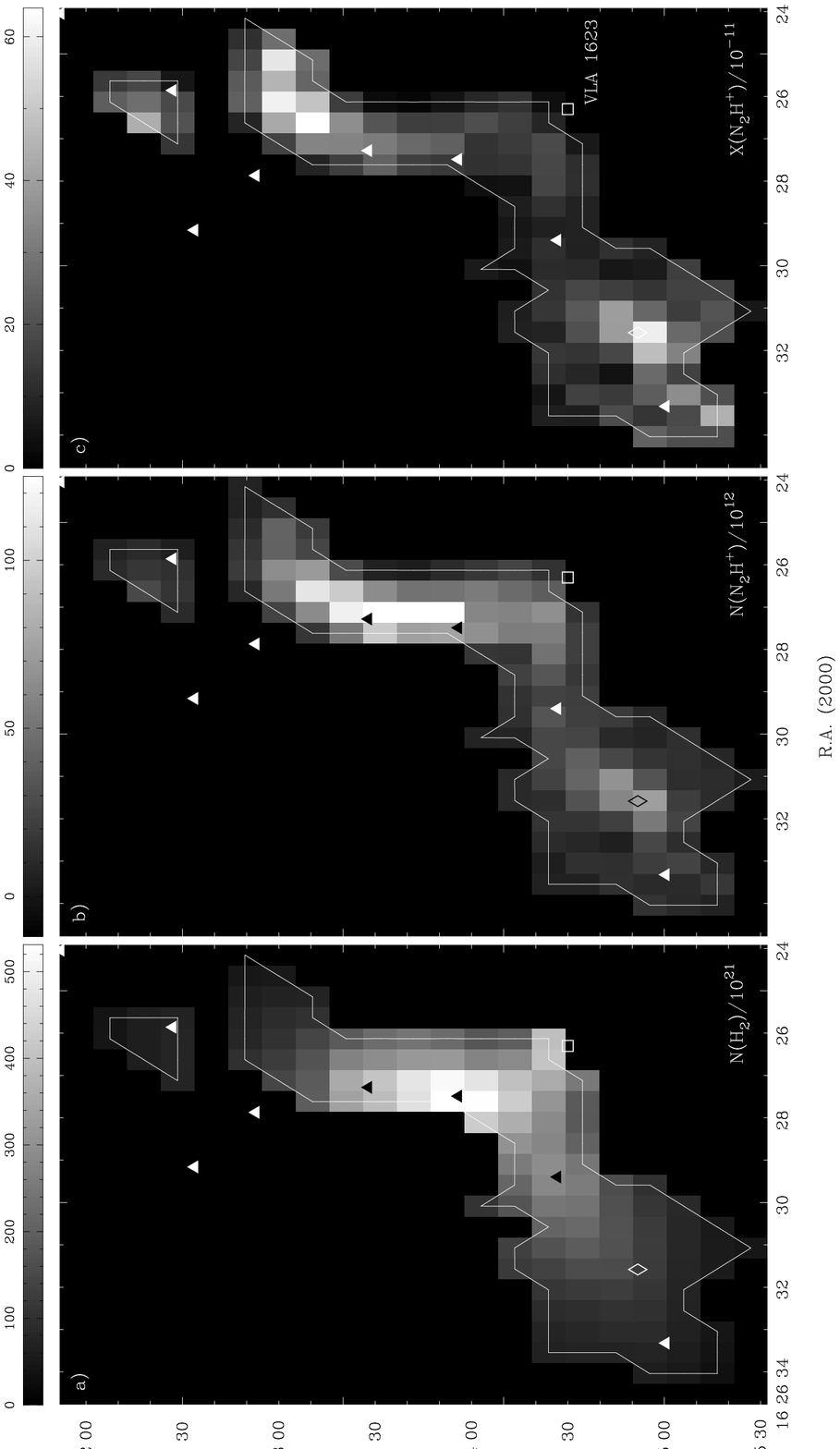}
\vspace{-2.25in}
\end{figure}

\begin{figure}
\vspace{7.25in}
\hspace{-0.25in}
\includegraphics{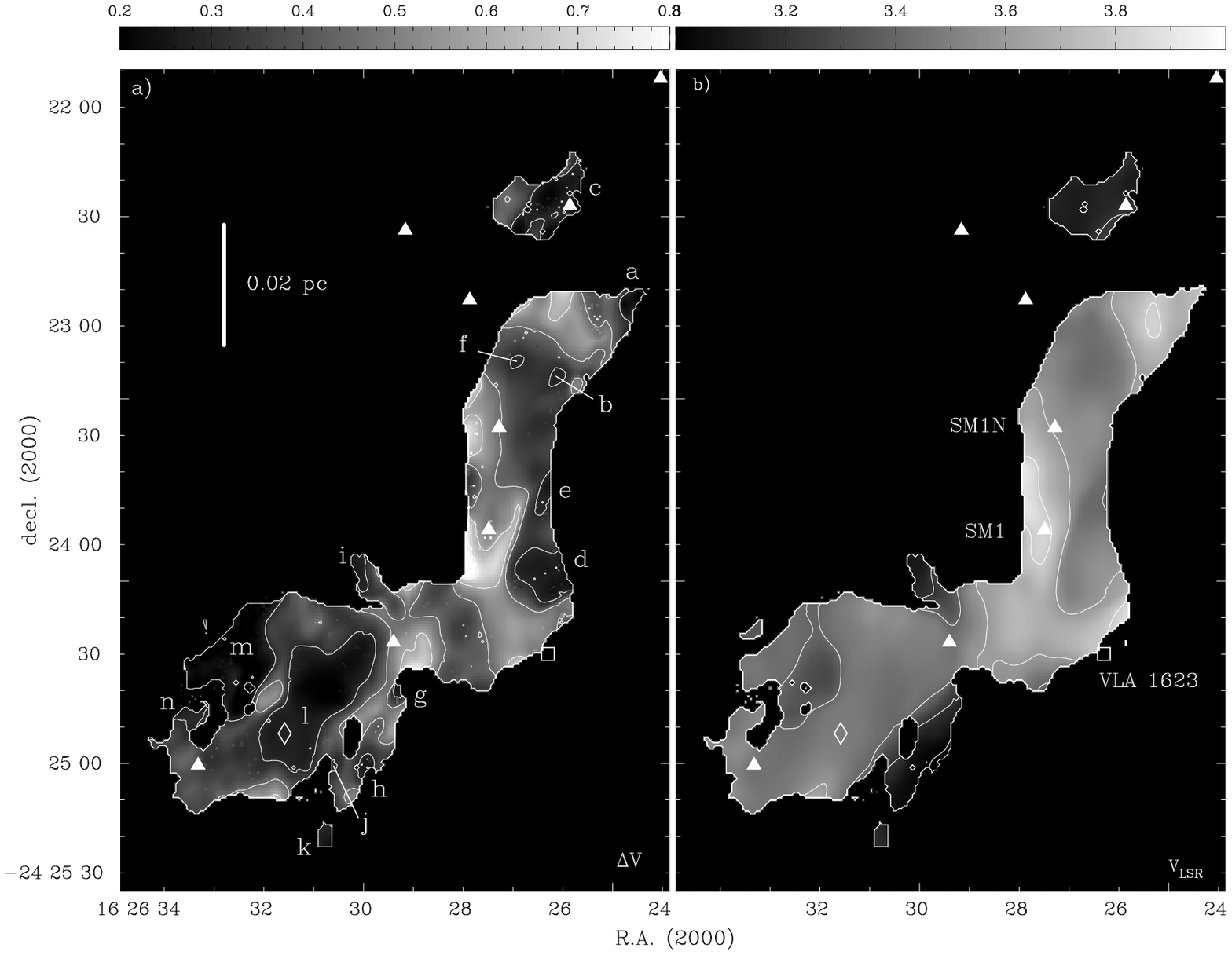}
\vspace{-2.25in}
\end{figure}

\begin{figure}
\vspace{7.25in}
\hspace{-0.25in}
\includegraphics{f4.ps}
\vspace{-2.25in}
\end{figure}

\begin{figure}
\vspace{7.25in}
\hspace{-0.25in}
\includegraphics{f5.ps}
\vspace{-2.25in}
\end{figure}

\begin{figure}
\vspace{7.25in}
\hspace{-0.25in}
\includegraphics{f6.ps}
\vspace{-2.25in}
\end{figure}

\begin{figure}
\vspace{7.25in}
\hspace{-0.25in}
\includegraphics{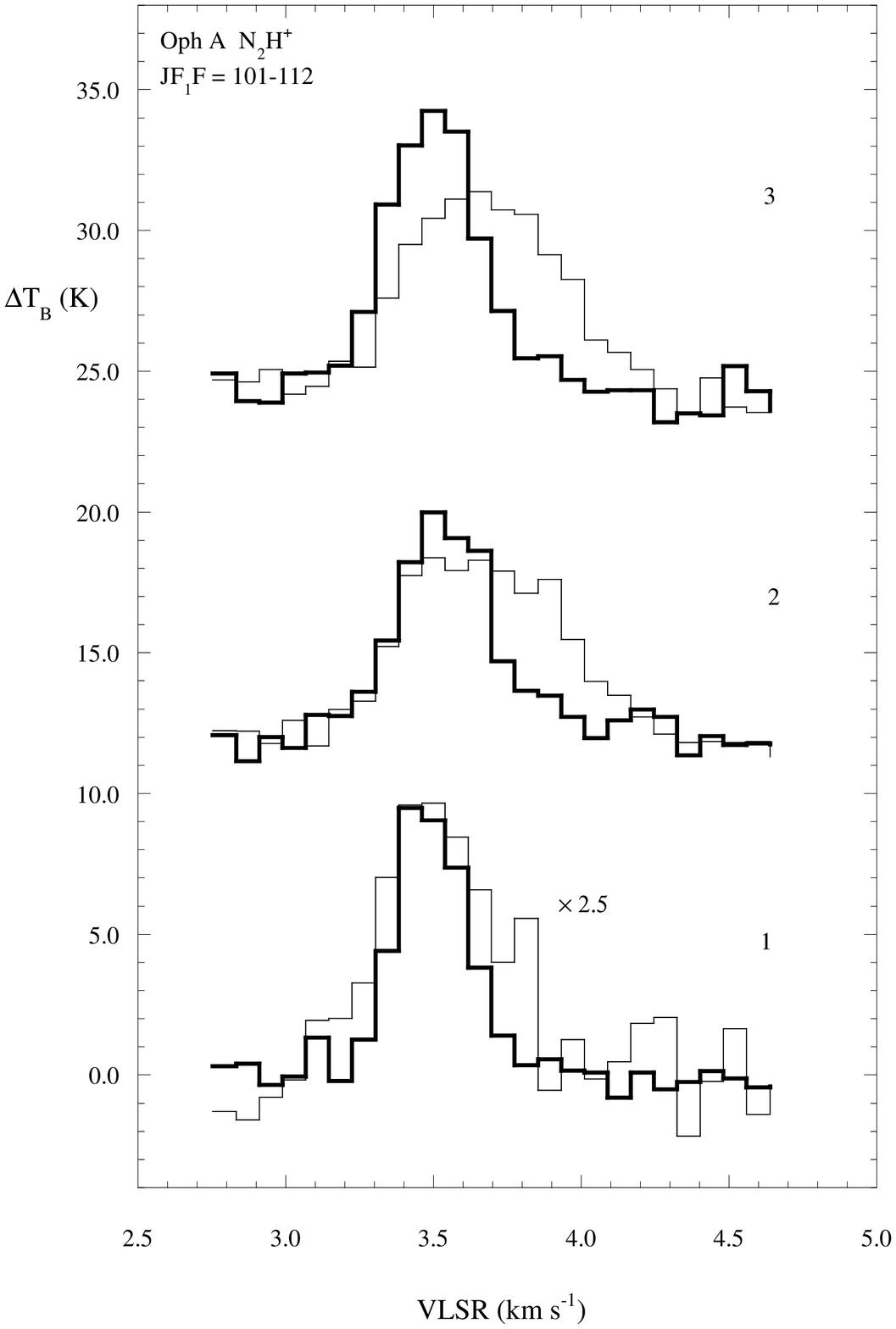}
\vspace{-2.25in}
\end{figure}

\end{document}